\documentclass[10pt,preprint]{emulateapj}

\usepackage{natbib}
\usepackage{rotate}
\def\gtorder{\mathrel{\raise.3ex\hbox{$>$}\mkern-14mu
             \lower0.6ex\hbox{$\sim$}}}
\def\ltorder{\mathrel{\raise.3ex\hbox{$<$}\mkern-14mu
             \lower0.6ex\hbox{$\sim$}}}

\newcommand{\feh}{\hbox{$ [{\rm Fe}/{\rm H}]$}}

\shorttitle{The old Galactic bulge in 3D from OGLE RR Lyrae stars}
\shortauthors{Pietrukowicz et al.}

\begin{document}

\title{Deciphering the 3D structure of the old Galactic bulge
from the OGLE RR Lyrae stars}

\author{P.~Pietrukowicz\altaffilmark{1},
S.~Koz{\l}owski\altaffilmark{1},
J.~Skowron\altaffilmark{1},
I.~Soszy{\'n}ski\altaffilmark{1},
A.~Udalski\altaffilmark{1},
R.~Poleski\altaffilmark{1,2},
{\L}.~Wyrzykowski\altaffilmark{1,3},
M.~K.~Szyma{\'n}ski\altaffilmark{1},
G.~Pietrzy{\'n}ski\altaffilmark{1,4},
K.~Ulaczyk\altaffilmark{1},
P.~Mr\'oz\altaffilmark{1},
D.~M.~Skowron\altaffilmark{1},
and M.~Kubiak\altaffilmark{1}
}

\altaffiltext{1}{Warsaw University Observatory,
Al. Ujazdowskie 4, 00-478 Warszawa, Poland}
\altaffiltext{2}{Department of Astronomy, Ohio State University,
140 W. 18th Ave., Columbus, OH 43210, USA}
\altaffiltext{3}{Institute of Astronomy, University of Cambridge,
Madingley Road, Cambridge CB3 0HA, UK}
\altaffiltext{4}{Universidad de Concepci{\'o}n, Departamento de F\'isica,
Casilla 160-C, Concepci{\'o}n, Chile}

\begin{abstract}
We have analyzed a sample of 27,258 fundamental-mode RR Lyrae variable
stars (type RRab) detected recently toward the Galactic bulge by
the Optical Gravitational Lensing Experiment (OGLE) survey.
The data support our earlier claim that these metal-poor stars trace
closely the barred structure formed of intermediate-age red clump giants.
The distance to the Galactic center (GC) inferred from the bulge RR Lyrae
stars is $R_0=8.27\pm0.01{\rm(stat)}\pm0.40{\rm(sys)}$~kpc. We show
that their spatial distribution has the shape of a triaxial ellipsoid
with an major axis located in the Galactic plane and inclined at an angle of
$i=20\degr\pm3\degr$ to the Sun--GC line of sight. The obtained scale-length
ratio of the major axis to the minor axis in the Galactic plane and to
the axis vertical to the plane is $1:0.49(2):0.39(2)$. We do not see the evidence
for the bulge RR Lyrae stars forming an X-shaped structure. Based on the
light curve parameters, we derive metallicities of the RRab variables and
show that there is a very mild but statistically significant radial metallicity
gradient. About 60\% of the bulge RRab stars form two very close sequences
on the period--amplitude (or Bailey) diagram, which we interpret as two
major old bulge populations: A and B. Their metallicities likely differ.
Population A is about four times less abundant than the slightly more
metal-poor population B. Most of the remaining stars seem to represent
other, even more metal-poor populations of the bulge. The presence
of multiple old populations indicates that the Milky Way bulge
was initially formed through mergers.
\end{abstract}



\section{Introduction}
\label{sec:intro}

RR Lyrae type variable stars are well-known tracers of old and metal-poor
populations. They play an essential role in the understanding of the
structure, formation, and evolution of the Milky Way. Multi-epoch
observations allow easy identification of this type of variable,
in particular of high-amplitude fundamental-mode RRab type stars with
characteristic tooth-shaped light curves. Based on a large number of observations
one can determine their properties with high precision, such as period,
metallicity, interstellar reddening, distance, and spatial distribution.

Thousands of RR Lyrae stars have been discovered in the Galactic
bulge, thick disk, and halo. Observations of the halo variables
(e.g. \citealt{2006AJ....132.1202K,2008ApJ...678..851K,
2010ApJ...708..717S,2012ApJ...756...23A,2012MNRAS.427.3374M,
2012MNRAS.424.2528S,2013ApJ...763...32D,2014ApJ...781...22Z,
2015MNRAS.446.2251T}) indicate that it probably consists of two
overlapping populations (\citealt{2008ApJ...678..865M,2009AcA....59..137S})
and is full of substructures from tidally disrupted dwarf galaxies
such as the Sagittarius dwarf spheroidal (Sgr dSph) galaxy
(\citealt{2001A&A...375..909C,2009AJ....137.4478K,2015MNRAS.446.2251T}).

Central regions of the Milky Way have been intensively monitored by the
Optical Gravitational Lensing Experiment (OGLE) since the early 1990s
(\citealt{1992AcA....42..253U}). Since 2010 March the project has been
in its fourth phase \citep[OGLE-IV,][]{2015AcA....65....1U}. With the
survey area increasing in each of the previous phases, the number of
detected RR Lyrae variables toward the Galactic bulge has increased
significantly: 215 stars in OGLE-I (e.g. \citealt{1997AcA....47....1U}),
2713 stars in OGLE-II (\citealt{2003AcA....53..307M}), and 16,836 stars
in OGLE-III (\citealt{2011AcA....61....1S}). The OGLE-III collection consisted
of 11,756 fundamental-mode (RRab), 4989 first-overtone (RRc), and 91
double-mode (RRd) stars. The analysis of that sample by
\cite{2012ApJ...750..169P} showed that the bulge RR Lyrae stars
seem to constitute a uniform population with a sharply peaked
metallicity distribution centered at $\feh=-1.02\pm0.18$~dex on the
\cite{1995AcA....45..653J} metallicity scale. A comparison of the
distributions of the dereddened brightness of RR Lyrae stars and red
clump (RC) giants indicated that in the inner regions ($|l|<3\degr$,
$|b|<4\degr$) the RR Lyrae stars tend to follow the barred structure
of RC stars inclined at an angle of about $30\degr$ with respect
to the line of sight between the Sun and the Galactic center (GC)
(\citealt{2007MNRAS.378.1064,2011A&A...534L..14G,2013MNRAS.434..595C,
2013ApJ...769...88N,2013MNRAS.435.1874W,2014AstL...40...86B}).

\cite{2013ApJ...776L..19D} combined $I$-band OGLE measurements
with the near-IR $K_s$-band data from the VISTA Variables in the
Via Lactea (VVV) survey (\citealt{2010NewA...15..433M}) for 7663 OGLE-III
RRab variables in order to refine the spatial distribution of the old bulge
population. These authors claim that the population of RR Lyrae stars has
a nearly spherical distribution, slightly elongated in its central part
almost toward us. They obtained a distance to the GC of
$R_0=8.33\pm0.05{\rm(stat)}\pm0.14{\rm(sys)}$~kpc and an
inclination angle of $12\fdg5\pm0\fdg5$.

Very recently, \cite{2014AcA....64..177S} presented a collection of 38,257
RR Lyrae variable stars detected in the OGLE fields covering over
182~deg$^2$ of the central regions of the Galactic bulge. In this set,
21,453 RR Lyrae stars are newly discovered objects in OGLE-IV.
The collection consists of 27,258 RRab, 10,825 RRc, and 174 RRd stars.
In this paper, we study the spatial structure of the old
Galactic bulge based on the whole available sample of the detected RRab
type stars from the OGLE survey. Variables of this type are on average
brighter, have higher amplitudes, and thanks to their characteristic
saw-tooth-shaped light curves are hard to overlook in comparison
to first-overtone sinusoidal-like RRc pulsators.

We also use the new large sample of RRab type variables to investigate
population properties of the Milky Way old bulge. In early studies,
classical bulges were assumed to be characterized by single stellar
populations whose stars formed on short timescales
(e.g. \citealt{1990ApJ...356..359H, 2000AJ....120..165T}).
However, modern spectroscopic studies have revealed a more complex picture
of star formation history via collapse, mergers, and secular processes
(e.g. \citealt{2006MNRAS.371..583,2008MNRAS.389..341M,2015MNRAS.446.2837S}).
RR Lyr stars as tracers of old populations may help to identify
the dominant mechanism responsible for the early formation of our Galaxy.

The outline of this paper is as follows. Section~\ref{sec:data} describes
the cleaning procedure of the sample. In Section~\ref{sec:sky}, we analyze
the observed distribution of the bulge RRab stars on the sky.
The analysis of the spatial structure is included in Section~\ref{sec:3d}.
In Section~\ref{sec:metal}, we investigate the photometric metallicity
gradient. In Section~\ref{sec:multipop}, we present the
discovery of multiple old populations among the bulge RR Lyrae stars.
Finally, Section~\ref{sec:conc} states the main conclusions of this work.


\section{Data sample}
\label{sec:data}

Prior to the analysis of the bulge RR Lyrae stars the original
sample from \cite{2014AcA....64..177S} required several cleaning steps.
From the whole sample of 27,258 RRab variables, we rejected 54 stars
as bona fide members and very likely members of eight globular
clusters (NGC 6441, NGC 6522, NGC 6540, NGC 6553, NGC 6558, NGC 6569, NGC 6642,
and NGC 6656). This procedure was based on \cite{2001AJ....122.2587C}
catalog of variable stars in Galactic globular clusters and its 2010
update.\footnote{http://www.astro.utoronto.ca/cclement/read.html}
From the original list, we also rejected object OGLE-BLG-RRLYR-02792
which was recently confirmed to be a low-mass binary component mimicking
an RR Lyrae pulsator (\citealt{2012Natur.484...75P,2013MNRAS.428.3034S}).

\begin{figure}
\centering
\includegraphics[width=8.5cm]{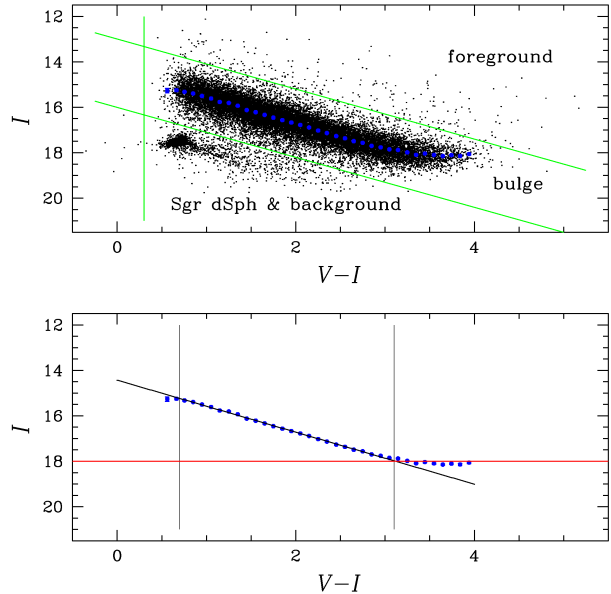}
\caption{Color--magnitude diagram (based on mean values) for all 27,258 RRab
variables observed in the OGLE fields (upper panel). Candidate bulge
variables are located inside the three green lines. Other stars
are very likely foreground and background objects. The lower sequence
is formed by variables from the Sgr dSph galaxy. Binning of the bulge data
every 0.1~mag in the $V-I$ color (blue points) shows that the bulge
sequence is practically linear for $0.7<V-I<3.1$~mag and complete down
to $I=18.0$~mag (lower panel).}
\label{fig:cmd}
\medskip
\end{figure}

In the next step, we cleaned the bulge sample from foreground and background
RR Lyrae stars by constructing the color-magnitude ($V-I$, $I$) diagram
(upper panel in Fig.~\ref{fig:cmd}). Due to different reddening toward observed
regions the bulge stars form a long sequence in the diagram.
By drawing three lines we roughly delimited bulge RR Lyrae variables from
background stars (line at $I=1.1(V-I)+16.0$), foreground stars (line
at $I=1.1(V-I)+13.0$), and stars with unreliable color (vertical line
at $V-I=0.3$~mag). Most of the background objects are variables from
the Sgr dSph galaxy. Other background variables and all foreground
variables very likely belong to the Milky Way thick disk and halo.
Binning of the bulge data shows that the sequence is clearly linear
for $0.7<V-I<3.1$~mag (lower panel in Fig.~\ref{fig:cmd}) and can be
described with the following relation:
\begin{equation}
\label{equ:1}
I=1.144(9)(V-I)+14.432(18),
\end{equation}
where its slope is the total-to-selective extinction $R_{I,V-I}=A_I/E(V-I)$.
It is 0.06 higher than the one found based on a sample from the OGLE-III
data that was a factor od two smaller (\citealt{2012ApJ...750..169P}).
The obtained value of $R_{I,V-I}$ is consistent with the fact that
the interstellar extinction toward the bulge is anomalous
(\citealt{2003ApJ...590..284U,2013ApJ...769...88N}).

The mean brightness of the bulge sequence clearly departs from linear
at $I\sim18$~mag or $V-I\sim3.1$~mag. Hence, we assume that our sample
of RRab stars is complete down to $I=18.0$~mag. The total number of
bulge variables with measured mean brightnesses in $V$ and $I$ amounts
to 21,026 objects.

\begin{figure}
\centering
\includegraphics[width=8.5cm]{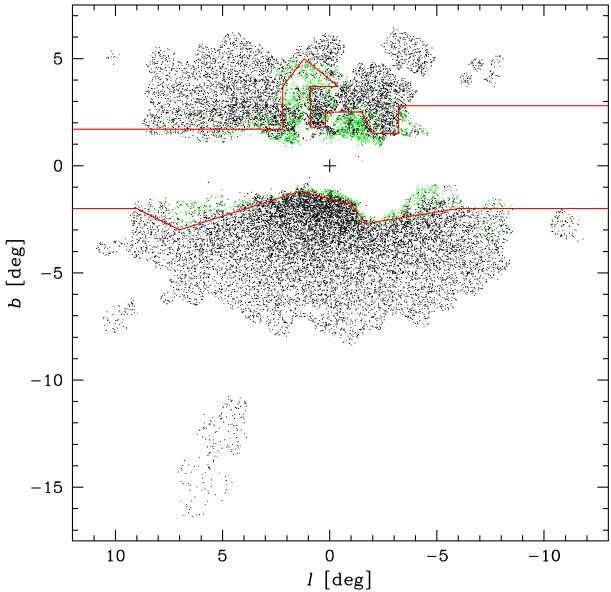}
\caption{Positions, in Galactic coordinates, of 21,026 OGLE bulge RRab
variables with measured brightnesses in both $V$ and $I$ band.
Black points are stars with a mean brightness of $I\leq18$~mag,
while green ones are those with a mean brightness of $I>18$~mag.
Red lines are limits between ''complete'' areas, toward which,
as we assume, all RRab stars have been detected (north of the upper
line and south of the lower one) and an ''incomplete'' stripe along
the Galactic plane due to high extinction.}
\label{fig:mapbulge}
\medskip
\end{figure}

In Fig.~\ref{fig:mapbulge}, we show a map with positions in Galactic
coordinates of all OGLE RR Lyrae variables with available $VI$ data
and distinguish between stars brighter and fainter than $I=18.0$~mag.
These stars concentrate toward the GC, but due to high
extinction they are not observed in the optical regime close to the
Galactic plane. By drawing two lines we set borders of two areas,
north and south of the plane, in which, as we assume, all bulge RRab stars
have been detected within the OGLE bulge fields. In our analysis
of the old bulge structure, we rely only on objects from the ``complete''
directions. The whole ``complete'' area amounts to 90.5~deg$^2$.

\section{Distribution on the sky}
\label{sec:sky}

Based on the OGLE-III data \cite{2012ApJ...750..169P} evidently confirmed
an earlier finding by \cite{1998IAUS..184..123M} that the bulge RR Lyrae population
is flattened along Galactic longitude. From the difference in longitudinal
and latitudinal surface density profiles, \cite{2012ApJ...750..169P} derived
an axis ratio of $\sim0.75$. For the present ``complete'' area covering 50\%
of the sky within an angular radius of $5\fdg5$ around the GC
and 28\% within $8\fdg75$, we can carry out a more accurate
determination of the observed on the sky shape of the old bulge.

\begin{figure}
\centering
\includegraphics[width=8.5cm]{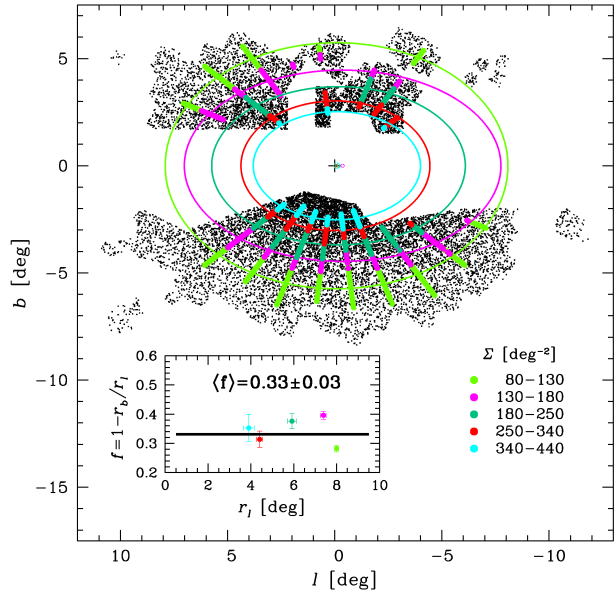}
\caption{Observed surface density of the bulge RRab stars determined
in ``complete'' directions up to $8\degr$ from the Galactic center.
Constant density lines are ellipses with the major axis along the
Galactic equator. Their centers are marked with small circles representing
the uncertainty. The observed semiminor to semimajor axis ratio
$r_b/r_l$ does not seem to change with the distance from the center.
The projected flattening $f=1-r_b/r_l$ has a mean value
of $0.33\pm0.03$ (inset).}
\label{fig:distrib}
\medskip
\end{figure}

We counted stars in cells in a radial coordinate system around $(l,b)=(0,0)$.
The cells were spaced every $15\degr$ in azimuth and they were
$0\fdg75$ long in radius. We sampled every $0\fdg05$ in radial
distance between $1\fdg25$ and $8\fdg0$ from the center.
The density in each cell was corrected for surface coverage.
Cells covering less then 80\% of the ``complete'' directions were
not taken into account. Results of the surface density determination
are presented in Fig.~\ref{fig:distrib}, where centers of the cells are
represented by points in five different colors. The colors code five
density ranges. To the points from the same density range we fitted ellipses
with the centers on the Galactic equator and described by the semimajor axis
$r_l$ and semiminor axis $r_b$. The obtained flattening $f=1-r_b/r_l$ in
a function of $r_l$ is plotted in the inset of Fig.~\ref{fig:distrib}.
It does not seem to change with the increasing distance, hence
we assume it to be constant at a mean value of $0.33\pm0.03$.

The most important implication from the observed surface density
distribution of the RR Lyrae stars is that the old bulge is not
spherically symmetric---it is ellipsoidal.

The uncertainty of the flattening $\sigma_f=0.03$ is much larger
than the effects of spherical projection. These effects are comparable
to the obtained $\sigma_f$ at a distance of about $30\degr$.
They are negligible at radial distances up to $10\degr$ from the center.

\begin{figure}
\centering
\includegraphics[width=8.5cm]{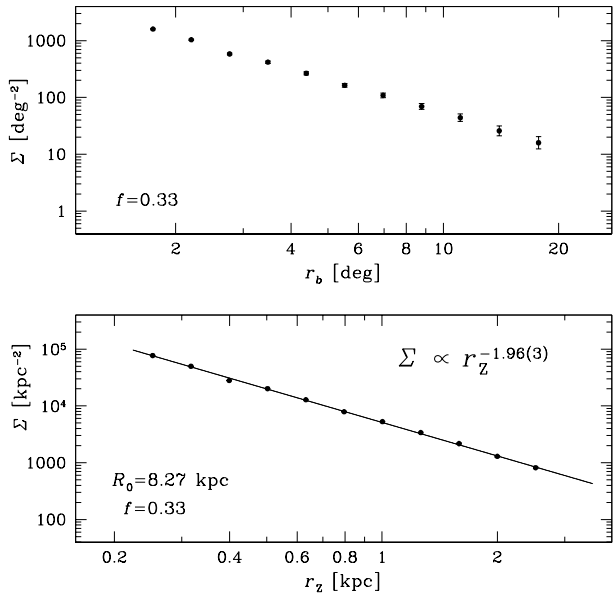}
\caption{
Radial surface density profile of RRab stars calculated for the
observed flattening $f$ and scaled to the minor axis in units of 
degrees (upper panel) and kpc (lower panel) assuming the distance
to the Galactic center $R_0$ determined in Section.~\ref{sec:3d}.
The obtained profile is a power law within the whole observed distance
range from the center.}
\label{fig:prof2D}
\medskip
\end{figure}

In two panels of Fig.~\ref{fig:prof2D}, we present the radial
surface density profile for the bulge RRab stars. We counted stars
in equally spaced annuli in a logarithmic angular distance scale from
the center at $(l,b)=(0,0)$. The data were corrected for completeness
and scaled to the minor (vertical) axis $r_b$ ($r_{\rm Z}$) by the factor
$r_b/r_l=0.67$. In the upper panel, we show the results in units of degrees,
while in the lower one, we convert the units of degrees to kiloparsecs
by adopting the distance to the GC obtained further
in Section~\ref{sec:3d}, $R_0=8.27$~kpc. The OGLE data span angular
distances between $1\fdg5$ and $19\fdg9$ or linear distances
from about 0.22~kpc to 2.8~kpc from the GC. As one can see,
the surface density profile of RRab stars in this Galactocentric
distance range can be represented by a single power law:
\begin{equation}
\label{equ:2}
{\rm log}~\Sigma = -1.96(3)~{\rm log}~r_{\rm Z} + 3.709(14).
\end{equation}
This result is much more precise than the one presented in
\cite{2012ApJ...750..169P} where the sky coverage and the RRab sample
were significantly smaller and variables were counted in OGLE-III
fields as units. Here we see that there is no break in the density
distribution at a distance of 0.5~kpc from the GC,
as reported in the earlier work.

\section{The 3D structure}
\label{sec:3d}

\subsection{Calculation of the distances to the RR Lyrae stars}

In this section, we derive parameters describing the spatial
structure of the RR Lyrae population. To obtain distances to individual
stars we have to correct their brightness for reddening. We calculated
dereddened mean magnitudes using
\begin{equation}
\label{equ:3}
I_0 = I - A_I,
\end{equation}
where the extinction $A_I$ was derived from the following formula
introduced in \cite{2013ApJ...769...88N}:
\begin{equation}
\label{equ:4}
A_I = 0.7465E(V-I) + 1.3700E(J-K).
\end{equation}
The reddening in the optical regime is
\begin{equation}
\label{equ:5}
E(V-I) = (V-I) - (V-I)_0,
\end{equation}
where the intrinsic color
\begin{equation}
\label{equ:6}
(V-I)_0 = M_V-M_I.
\end{equation}
In Equation~(\ref{equ:6}), the absolute brightnesses $M_V$ and $M_I$ were
computed from the theoretical relations given in \cite{2004ApJS..154..633C}:
\begin{equation}
\label{equ:7}
M_V = 2.288 + 0.882~\log Z + 0.108~(\log Z)^2,
\end{equation}
\begin{equation}
\label{equ:8}
M_I = 0.471 - 1.132~\log P + 0.205~\log Z,
\end{equation}
with the following conversion for metallicity:
\begin{equation}
\label{equ:9}
{\rm log} Z = {\rm [Fe/H]} - 1.765.
\end{equation}
The zero points of the relations for $M_V$ and $M_I$ were calibrated
to the data obtained for the well-studied representative globular cluster
M3 (NGC 5272, \citealt{2004ApJ...600..409C}), which hosts about 240 RR
Lyrae variables (\citealt{2001AJ....122.2587C}).
We measured the metallicities on the \cite{1995AcA....45..653J} scale
(J95) by applying an empirical relation from \cite{2005AcA....55...59S}:
\begin{equation}
\label{equ:10}
{\rm \feh_{J95}} = - 3.142 - 4.902 P + 0.824 \phi_{31},
\end{equation}
where $P$ is the pulsation period and $\phi_{31}=\phi_3-3\phi_1$
is a Fourier phase combination for sine decomposition of the
$I$-band light curve. The reddening $E(J-K)$ was taken from the maps
in \cite{2012A&A...543A..13G} which were prepared in the
framework of the VVV survey (\citealt{2010NewA...15..433M}).
It is available for RR Lyr stars located within the VVV bulge
area of the following limits: $-10\fdg1 \leq l \leq10\fdg4$
and $-10\fdg4 \leq b \leq +5\fdg1$.

Finally, the distance $d$ to each RRab stars was calculated from
the equation
\begin{equation}
\label{equ:11}
d = 10^{1 + 0.2(I_0-M_I)}.
\end{equation}

\subsection{Distance to the GC}

\begin{figure}
\centering
\includegraphics[width=8.5cm]{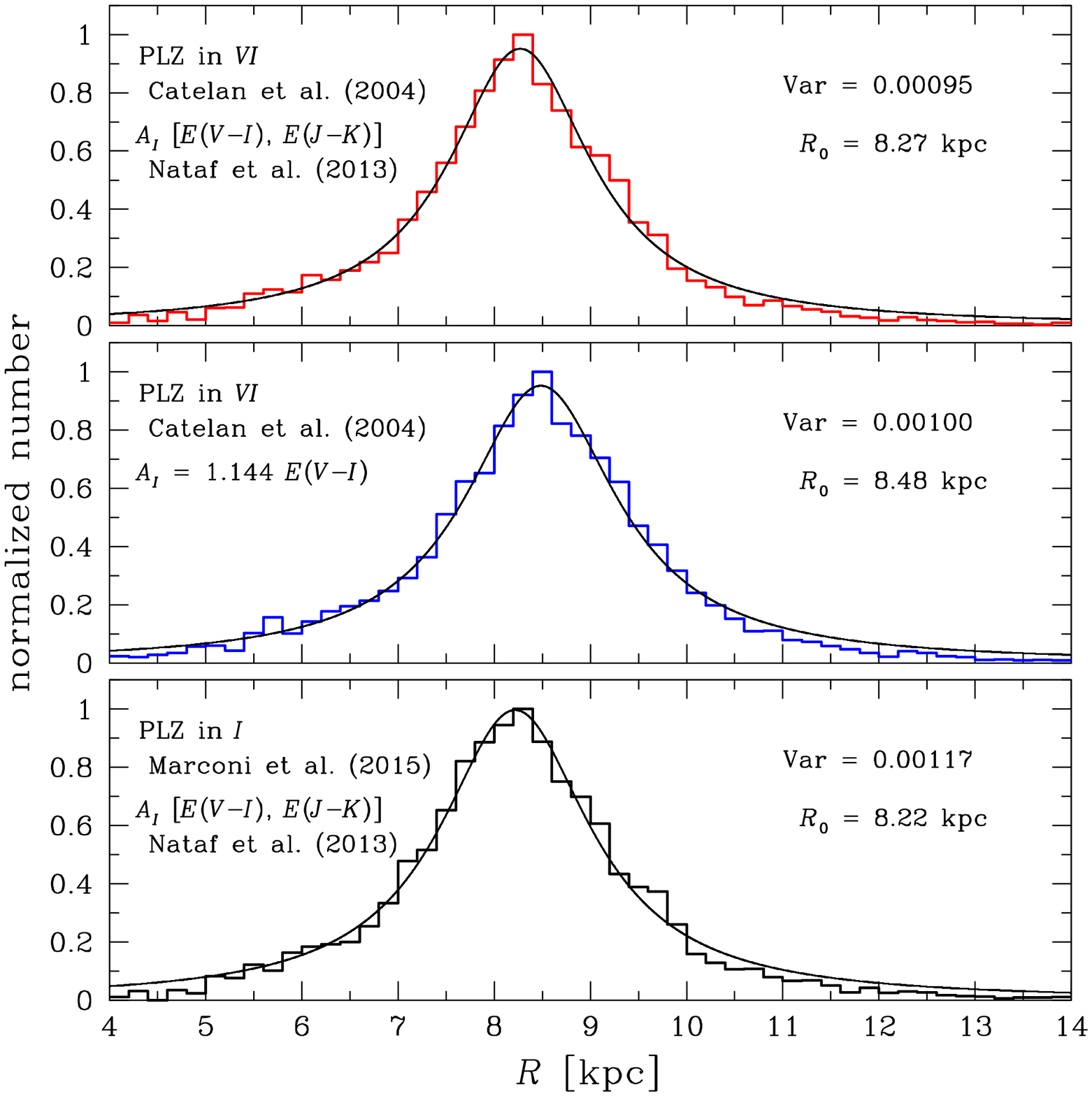}
\caption{Distance distribution for RRab stars from an area with
Galactic coordinates $-4\degr<l<+4\degr$ and $-6\fdg7<b<-2\fdg7$
obtained with three different methods: using theoretical
period--luminosity--metallicity (PLZ) relations in the $VI$ bands
from \cite{2004ApJS..154..633C} and \cite{2013ApJ...769...88N}
relation for extinction with optical and near-IR components (upper panel),
using theoretical relations also from \cite{2004ApJS..154..633C}
but with a simple linear correction for absorption with the
total-to-selective extinction $R_{I,V-I}=1.144$ (middle panel),
and by applying a theoretical period--luminosity--metallicity
relation in the $I$ band from \cite{2015ApJ...808...50M} and the
formula for extinction from \cite{2013ApJ...769...88N} (lower panel).
The first method gives the smallest variance of residuals and a maximum
value at $R_0=8.27$~kpc, which is the closest to the most recent
estimates of the distance to the Galactic center of about 8.3~kpc.
The third method leads independently to very similar distance estimates.}
\label{fig:histdist}
\medskip
\end{figure}

Having distances to individual RR Lyrae stars we can estimate the
distance to the center of the entire old population, which we assume
to coincide with the center of our Galaxy. Inside the ``complete''
OGLE directions, we selected a rectangle area symmetric in Galactic
longitude with $-4\degr<l<+4\degr$ and $-6\fdg7<b<-2\fdg7$.
In the upper panel of Fig.~\ref{fig:histdist}, we plot the distribution
of the distances to 4689 stars from this area calculated using
the relation on extinction from \cite{2013ApJ...769...88N}.
To the distribution we fit a \cite{1962AJ.....67..471K} -like profile:
\begin{equation}
\label{equ:12}
n=\frac{n_0}{1+\left(\frac{d-R_0}{r_c}\right)^2},
\end{equation}
where $n_0$ is the maximum number of stars at the distance $R_0$,
and $r_c$ is the core radius. It is important to note that
prior to the fitting we applied two geometric corrections to the data.
First, the individual distances were projected onto the Galactic plane
(by ${\rm cos}~b$). Second, we scaled the distribution by $d^{-2}$ to
correct for the so-called cone effect where more objects are seen at
larger distances inside a solid angle. The maximum of the distribution
is at a distance $R_0=8.27$~kpc. The obtained value is close
to the most recent determinations of the distance to the GC:
$8.28\pm0.15({\rm stat})\pm0.29({\rm sys})$~kpc (\citealt{2009ApJ...707L.114G}),
$8.27\pm0.29$~kpc (\citealt{2012MNRAS.427..274S}),
$8.33\pm0.05({\rm stat})\pm0.14({\rm sys})$~kpc (\citealt{2013ApJ...776L..19D}),
$8.34\pm0.16$~kpc (\citealt{2014ApJ...783..130R}),
$8.27\pm0.09({\rm stat})\pm0.1({\rm sys})$~kpc (\citealt{2015MNRAS.447..948C}).

For a comparison, in the remaining two panels of Fig.~\ref{fig:histdist},
we show distance distributions obtained in two other ways.
In the middle panel, we apply a simple correction for reddening
by taking the mean value of $R_{I,V-I}=1.144$ determined
earlier in Section~\ref{sec:data}---as was done
in \cite{2012ApJ...750..169P}. In the lower panel, we calculate
the distances through a period--luminosity--metallicity relation
in the $I$ band for RRab stars provided by
\cite{2015ApJ...808...50M} in their Table~6:
\begin{equation}
\label{equ:13}
M_I = - 0.07 - 1.66 {\rm log} P + 0.17 {\rm \feh_{C09}},
\end{equation}
where the metallicity is on a scale for globular clusters
\citep[][C09]{2009A&A...508..695C}. In this approach, we transformed the
metallicities of bulge variables from the J95 scale to C09 via the
\cite{1984ApJS...55...45Z} scale (ZW84) using a relation from
\cite{2000AJ....119..851P}:
\begin{equation}
\label{equ:14}
{\rm \feh_{ZW84}} = 1.028 {\rm \feh_{J95}} - 0.242
\end{equation}
and the following formula from \cite{2009A&A...508..695C}:
\begin{equation}
\label{equ:15}
{\rm \feh_{C09}} = 1.105 {\rm \feh_{ZW84}} + 0.160.
\end{equation}
Here we also use \cite{2013ApJ...769...88N} formula for extinction
with $E(J-K)$ from VVV maps in \cite{2012A&A...543A..13G} and $E(V-I)$
from OGLE-III maps in \cite{2013ApJ...769...88N}, in both cases determined
from RC giants. As one can see from Fig.~\ref{fig:histdist}, the method
based on theoretical relations from \cite{2004ApJS..154..633C} and the
formula for extinction from \cite{2013ApJ...769...88N} gives the lowest
variance of residuals of the fit and $R_0$ closest to the most recent
estimates of the distance to the Galactic center. The approach with
a non-standard but constant-in-all-directions value of $R_{I,V-I}$
gives a variance of the residuals that is higher by about
5\% and a larger value of $R_0$. The third method,
which relies on an independent set of stellar models,
is in a very good agreement with our first approach. This shows that
both methods are solid. If the reddening maps covered the whole
OGLE-IV bulge area, the approach via \cite{2015ApJ...808...50M}
PLZ relation could be used for the whole set of RRab stars.

The statistical uncertainty of the distance to the GC
based on thousands of RR Lyrae stars is expected to be small.
If we adopt a mean accuracy of OGLE brightness measurements at a level
of $\sigma_I=\sigma_V=0.02$~mag (\citealt{2008AcA....58...69U}),
an accuracy of the derived metallicity of 0.005~dex at $\feh_{\rm J95}=-1.02$~dex,
and a mean accuracy of the reddening $E(J-K)$ of 0.104~mag, we find
a mean uncertainty of the distance modulus to the individual stars
of 0.147~mag, and hence an uncertainty of the distance of $\sigma=0.56$~kpc.
The selected rectangle area contains $n=4689$ variables, which
gives a final statistical error of the distance to the GC
$\sigma_{\rm R0,stat}=\sigma/\sqrt{n}=0.008$~kpc.

The systematic error of $R_0$ relies on several estimations. In particular,
it relies on the quality of estimation of the absolute brightness $M_V$
derived from the luminosity--metallicity dependence. It does not exceed
$\Delta M_V=0.1$~mag if one compares quadratic formulae from \cite{2004ApJS..154..633C},
\cite{2006ARA&A..44...93S}, and \cite{2007A&A...476..779B} for metallicities
around $\feh_{\rm J95}=-1.0$~dex. The relation for extinction $A_I$ (Eq.~\ref{equ:4})
is based on RC giants for which the intrinsic colors are known to an accuracy
of $\Delta (V-I)_{\rm RC,0}=0.01$~mag and $\Delta (J-K)_{\rm RC,0}=0.02$~mag
(\citealt{2012A&A...543A..13G,2013ApJ...769...88N}). This gives an error of
$\Delta A_I=0.028$~mag. The uncertainty of the metallicity relation
(Equation~(\ref{equ:9})) of 0.18~dex (\citealt{2005AcA....55...59S})
propagates to a small additional error of 0.014~mag in the $(V-I)_0$ color
of RR Lyrae stars. If we take into account all of the above errors
in the equation for the distance modulus:
\begin{equation}
\label{equ:16}
{\mu}=I-A_I-M_V+(V-I)_0,
\end{equation}
we find $\sigma_{\mu,sys}=0.105$~mag, and hence the systematic uncertainty
of the distance to GC of $\sigma_{\rm R0,sys}\sim0.40$~kpc.
Finally, we have $R_0=8.27\pm0.008({\rm stat})\pm0.40({\rm sys})$~kpc.
As expected, the uncertainty of $R_0$ is dominated by systematic error, whereas
the large number of variables significantly lowers the statistical one.

\subsection{Geometry of the old bulge}

\begin{figure}
\centering
\includegraphics[width=8.5cm]{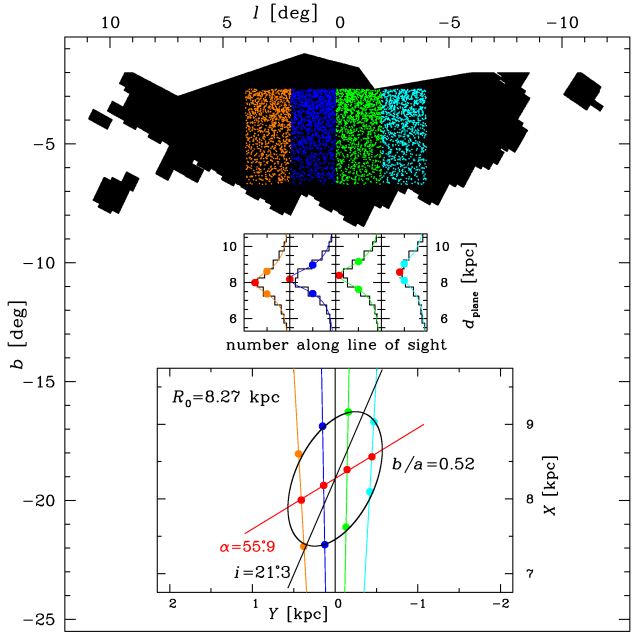}
\caption{Results from the analysis of the number density distribution
of bulge RRab stars along four selected directions. The centers of
the four analyzed rectangular areas, shown with different colors, have
the same Galactic latitude of $b=-4\fdg7$ and are equally spaced
in Galactic longitude from $l=-3\degr$ to $l=+3\degr$. The inset in the
middle presents the obtained density distributions. The maxima of the
distributions, marked with red dots, clearly change with the Galactic
longitude. Four pairs of dots represent the same density level
in the four investigated directions. In the projection onto the Galactic
plane (lower inset), they form an ellipse tilted at an angle
of $i\approx21\degr$ to the Sun--GC line of sight and the axis ratio
$b/a\approx0.52$. A value of $R_0=8.27$~kpc was applied.
The Sun is located at $(X,Y)=(0,0)$.}
\label{fig:polarpro}
\medskip
\end{figure}

\begin{figure}
\centering
\includegraphics[width=8.5cm]{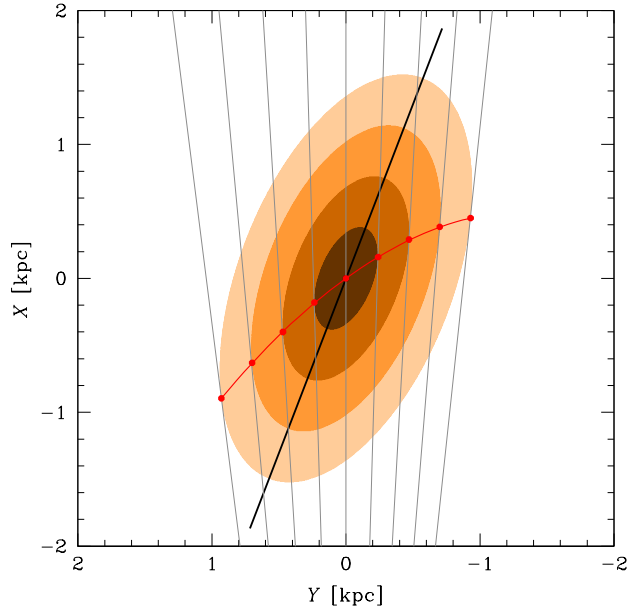}
\caption{Central part of the Galactic bulge, presented schematically,
as it would be seen from the Galactic pole (the one at the positive
Galactic latitude). We generate four constant density ellipses with
$b/a=0.5$ inclined at $21\degr$ to the line of sight of the observer
located at $(Y,X)=(0,-8.27)$~kpc. The black thick line shows the real
orientation of the bar's long axis. The tangential lines (in gray)
to the ellipses, starting from the observer, show the points of
the highest stellar density along these observer lines of sight
(red points). The highest density points are often incorrectly
interpreted as coinciding with the long axis of the bar, resulting
in a wrong, overestimated angle of the bar (see red vs. black lines
in the lower inset in Fig.~\ref{fig:polarpro}).}
\label{fig:bar}
\medskip
\end{figure}

To find the shape of the old bulge we use stars from the same rectangular
area. We divide the area into four parts equal in longitude as shown with
different colors in the upper part of Fig.~\ref{fig:polarpro}. The parts
from lower to higher longitude (from right to left) contain 1060, 1292,
1295, and 1042 objects, respectively. For each part, separately, we plot the
distance distribution and fit the King-like profile, as presented in the middle
part of Fig.~\ref{fig:polarpro}. The colors remain the same for each of the four
probed sightlines, associated with the weighted centers of the divided parts.
We mark the maxima of the fits with red dots. It is clearly seen that
the distance to the maximum decreases with the increasing Galactic
longitude. In the projection onto the Galactic plane, the maxima form
a line tilted at a relatively high angle of about $56\degr$
with respect to the Sun-GC line, as it is illustrated
in the lower part of Fig.~\ref{fig:polarpro}. The line for the highest
density is not straight but slightly curved, as shown in an auxiliary
figure, Fig.~\ref{fig:bar}, and cannot be interpreted as the bar's long axis.
To find the true shape and inclination angle of the old bulge,
along the four sightlines we mark positions at the same density level,
here set at half maximum of the distribution toward $(l,b)=(-1\degr,-4\fdg7)$.
As one can see, such points form an ellipse with an axis ratio of $b/a\approx0.52$
tilted at an angle of $i\approx21\degr$ to the Sun-GC line. This demonstrates
that the old bulge traced by the RR Lyrae stars is a triaxial
ellipsoid with the major axis $a$ and minor axis $b$ located in the
Galactic plane and the axis $c$ perpendicular to it.

\begin{figure}
\centering
\includegraphics[width=8.5cm]{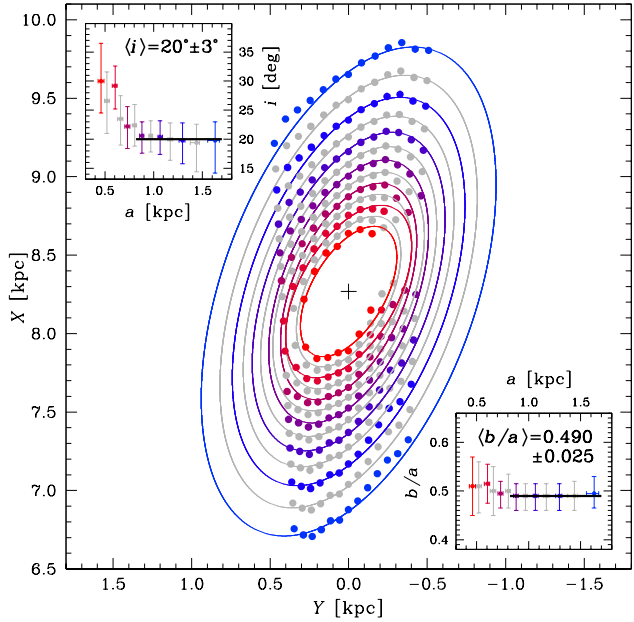}
\caption{Elliptical density contours fitted to the data along
13 sightlines, from $l=-3\degr$ to $l=+3\degr$ every $\Delta l=0\fdg5$
and projected onto the Galactic plane. The Sun is located at $(X,Y)=(0,0)$.
For clarity, every other density level is shown in gray.
The insets present the obtained inclination angle $i$ (upper left)
and axis ratio $b/a$ (lower right) as a function of the semimajor
axis $a$ of the ellipse. Ellipses fitted to points at distances
0.9--1.7~kpc have almost the same inclination angle and axis ratio:
$i\approx20\degr$ and $b/a\approx0.49$. Closer to the center, our results
are likely biased by the size of the data bin ($2\degr$ in longitude)
which is comparable to the size of the ellipse. Both parameters
of the ellipse probably stay the same.}
\label{fig:dencont}
\medskip
\end{figure}

In the next step, we probe the same area in the sky but for
more directions and a wider range of density levels. We selected
13 directions between Galactic longitudes $l=-3\fdg0$
and $l=+3\fdg0$ every $0\fdg5$ at the same latitude $b=-4\fdg7$.
The data were taken in a rectangular box of the same size,
$2\degr$ wide in longitude and $4\degr$ high in latitude,
and projected onto the Galactic plane, as it was done above.
To each of 11 density levels we fit an ellipse with the center at
$(X,Y)=(8.27, 0)$~kpc and described by the semi-major axis $a$,
axis ratio $b/a$, and inclination $i$. Results of the $\chi^2$
minimization of these three parameters are presented
in Fig.~\ref{fig:dencont}. Both the inclination angle and axis
ratio are practically constant in the range of probed distances
from the center between 0.9~kpc and 1.7~kpc. We find 
$i=20\degr\pm3\degr$ and $b/a=0.490\pm0.025$. Closer to the
center we obtain different values. In particular, the tilt angle
increases up to about $30\degr$. We suspect that the results for the
inner part are affected an ellipse of similar size to that
used to bin data. The angle and axis ratio probably stay the same
toward the center. However, we cannot rule out a possibility that
they are different.

Knowing the observed flattening on the sky $f$, axis ratio $b/a$,
and inclination angle $i$, we can estimate the axis ratio $c/a$
of the old bulge ellipsoid (see the appendix). For $f=0.33\pm0.03$,
$b/a=0.490\pm0.025$, and $i=20\degr\pm3\degr$ we find
$c/a=0.384\pm0.015$ with the uncertainty returned from Monte Carlo
simulations.

At the end of Section~\ref{sec:sky}, we showed that for RRab variables
the surface density profile along the vertical axis $r_{\rm Z}$ can be described
as a single power law. Since the positions on the sky are much more
accurate than distances to the stars and the $Z$ axis is the true minor axis
of the ellipsoid, from the surface density profile we can easily find
the spatial density distribution as
\begin{equation}
\label{equ:17}
{\rm log}~\rho = -2.96(3)~{\rm log}~r_{\rm Z} + 1.8545(7),
\end{equation}
with $\rho$ in kpc$^{-3}$. The obtained power-law index of $-2.96\pm0.03$
is comparable to the value of $-2.78\pm0.02$ reported for the inner halo
RR Lyrae variables in \cite{2012ApJ...756...23A} and $-2.8$ for halo main
sequence stars near the turn-off point determined by \cite{2008ApJ...673..864J}.
This result supports an idea that the old bulge and halo form
a uniform Galactic component.

\subsection{Absence of an X-shaped structure in RR Lyrae stars}

\begin{figure}
\centering
\includegraphics[width=8.5cm]{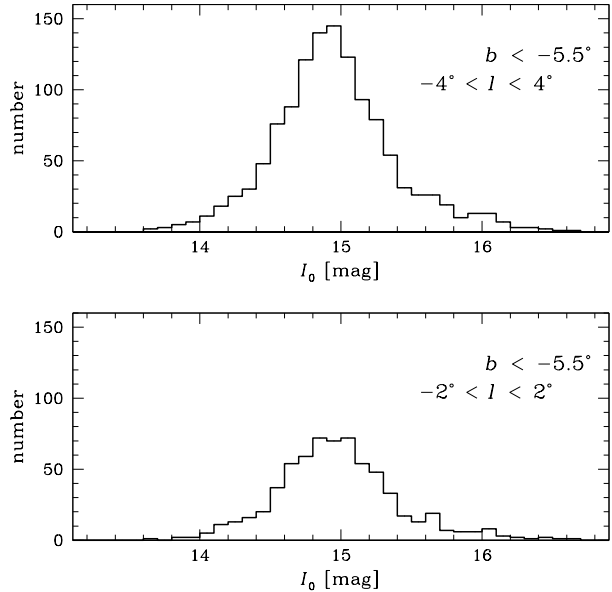}
\caption{
Distributions of the dereddened brightness $I_0$ for OGLE RRab stars in
regions where intermediate-age red clump giants form a prominent X-shaped
structure of the Galactic bar. Both distributions are for stars with
Galactic latitudes $b<-5\fdg5$ and around $l=0\degr$, but the top
distribution is for a range in longitudes that is twice as wide
as the bottom one. There is no split in the old bulge traced by
RR Lyrae stars.}
\label{fig:noxshape}
\medskip
\end{figure}

With the present sample of OGLE RR Lyrae variables, we can verify whether
they follow the same spatial structure in regions where the bulge RC
giants show a split in brightness (\citealt{2010ApJ...721L..28N})
and form an X-shaped bulge (\citealt{2010ApJ...724.1491M}).
The difference in the $I$ band for RC stars reaches about 0.5~mag at
Galactic latitudes $|b|\sim5\fdg5$. In Fig.~\ref{fig:noxshape}, we plot
histograms of dereddened $I$-band brightness derived for two sets
of RRab stars, both for objects with latitudes $-8\fdg5<b<-5\fdg5$
but with different longitude ranges around $l=0\degr$. No split is
observed. This confirms that only metal-rich bulge populations
have this feature (\citealt{2012ApJ...756...22N,2012A&A...546A..57U}).

\section{Radial metallicity gradient}
\label{sec:metal}

\begin{figure}
\centering
\includegraphics[width=8.5cm]{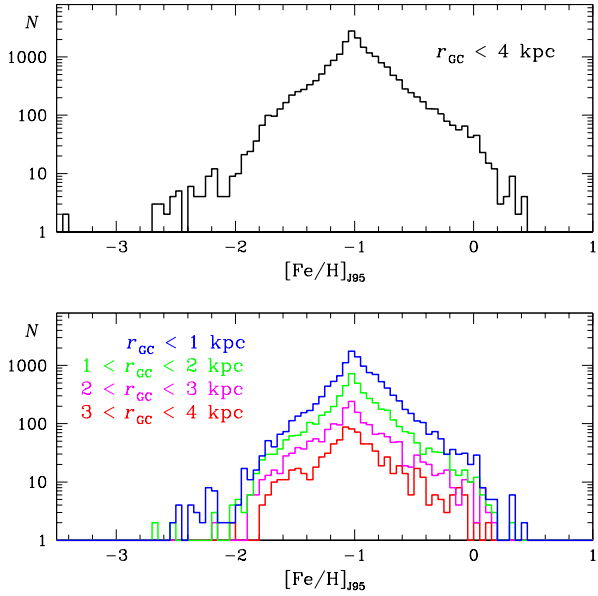}
\caption{
Metallicity distribution on the \cite{1995AcA....45..653J} scale
for all bulge RRab variables with measured Galactocentric distances
$r_{\rm GC}<4.0$~kpc (top) and for four separate distance ranges (bottom).
With increasing distance, the maximum of the distribution
slightly moves toward lower metallicities but its shape stays the same.}
\label{fig:histfeh}
\medskip
\end{figure}

\begin{figure}
\centering
\includegraphics[width=8.5cm]{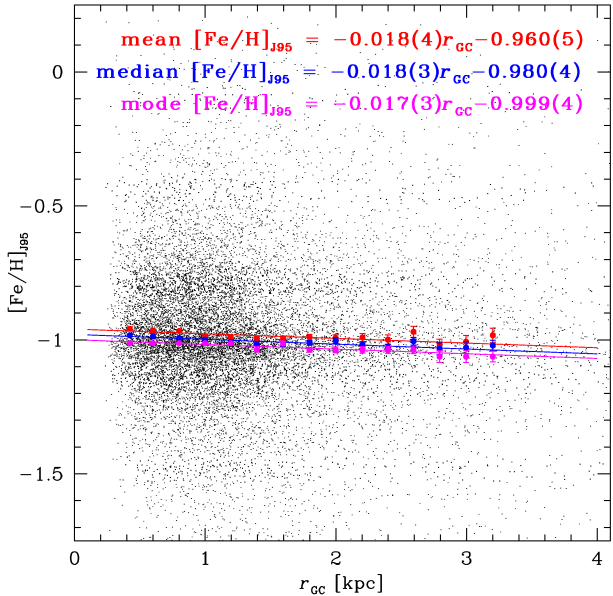}
\caption{
Metallicity as a function of distance from the Galactic center
for RRab variables. Relations for average, median, and modal values
calculated in 0.2-kpc bins are presented with different colors.
The decrease of the metallicity with the distance is very mild but
significant ($>4\sigma$). Almost the same slope for all three relations
indicates no changes in the metallicity distribution with the distance
from the center.}
\label{fig:fehrad}
\medskip
\end{figure}

The well-sampled OGLE light curves of RRab variables allow us to assess
metallicities of the stars using the method proposed by
\cite{1995A&A...293L..57K} and further developed by \cite{1996A&A...312..111J}.
The iron abundances, which we already derived in the previous section
for about 98.5\% of stars with a given $\phi_{31}$ value, are on the
\cite{1995AcA....45..653J} scale. In the top panel of Fig.~\ref{fig:histfeh},
we show the metallicity distribution for 19,869 bulge RRab stars
with a measured Galactocentric distance of $r_{\rm GC}<4.0$~kpc by binning the
data every 0.05~dex. The distribution has the peak at $\feh_{\rm J95}=-1.02$~dex
and a dispersion of 0.25~dex. The same values were reported in
\cite{2012ApJ...750..169P} for a sample of 10,259 OGLE-III RRab stars.
A much larger sample of stars analyzed here allows for verification of
a radial metallicity gradient recently noticed by \cite{2014A&A...571A..59S}.
They found it to be very small but relatively significant:
$-0.016\pm0.008$~dex~kpc$^{-1}$.
In Fig.~\ref{fig:fehrad}, we plot the metallicity of bulge RRab stars as
a function of distance from the GC. We fit linear regression separately
to the mean, median, and modal values determined in 0.2~kpc wide
bins from 0.3~kpc to 3.3~kpc. The errors of the bin values are weighted
by the square root of the number of stars in each bin. In all three cases
the obtained gradient is indeed mildly negative but statistically
significant ($>4\sigma$). Almost the same slopes for all three measures
indicate no change in the shape of the metallicity distribution
with distance. A delicate asymmetry of the distribution toward
lower metallicities is confirmed in the inequality mode$<$median$<$mean
at any distance from the center. Constant differences between the three
values indicate no changes in the shape of the distribution with the distance.
This can also be noted in the bottom panel of Fig.~\ref{fig:histfeh}.

\section{Multiple old populations}
\label{sec:multipop}

\begin{figure}
\centering
\includegraphics[width=8.5cm]{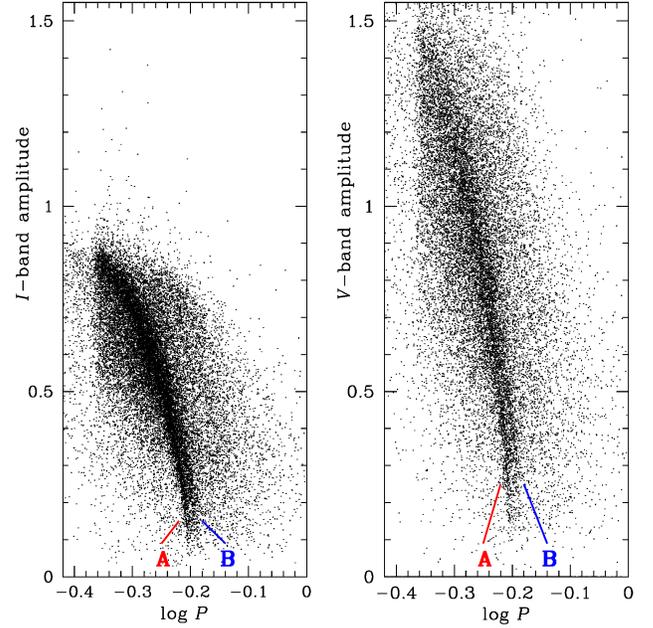}
\caption{
Period--amplitude diagrams for OGLE bulge RRab variables in two observed
bands. Two distinct sequences, A and B, are well visible, in particular
for low-amplitude stars.}
\label{fig:pampvi}
\medskip
\end{figure}

\begin{figure}
\centering
\includegraphics[width=8.5cm]{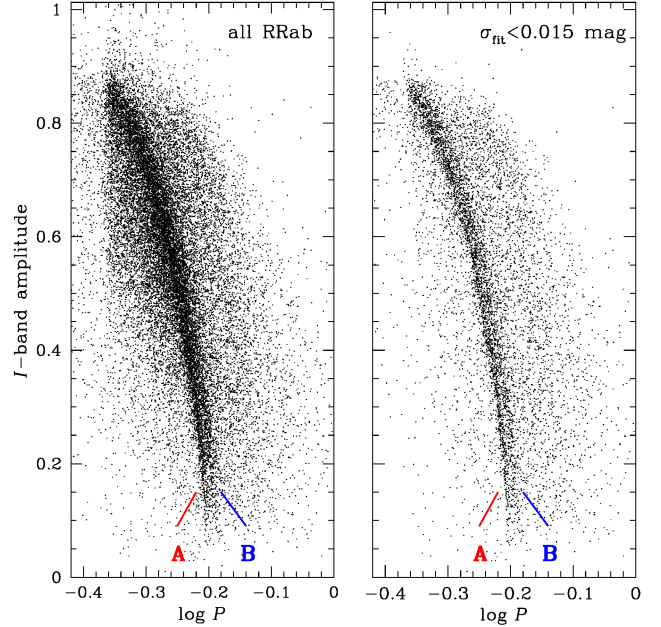}
\caption{
Period--amplitude diagram in the $I$ band for all bulge RRab
variables (left panel) and objects with the scatter around the fit
$\sigma_{\rm fit}<0.015$~mag (right panel).}
\label{fig:pamp015}
\medskip
\end{figure}

\begin{figure}
\centering
\includegraphics[width=8.5cm]{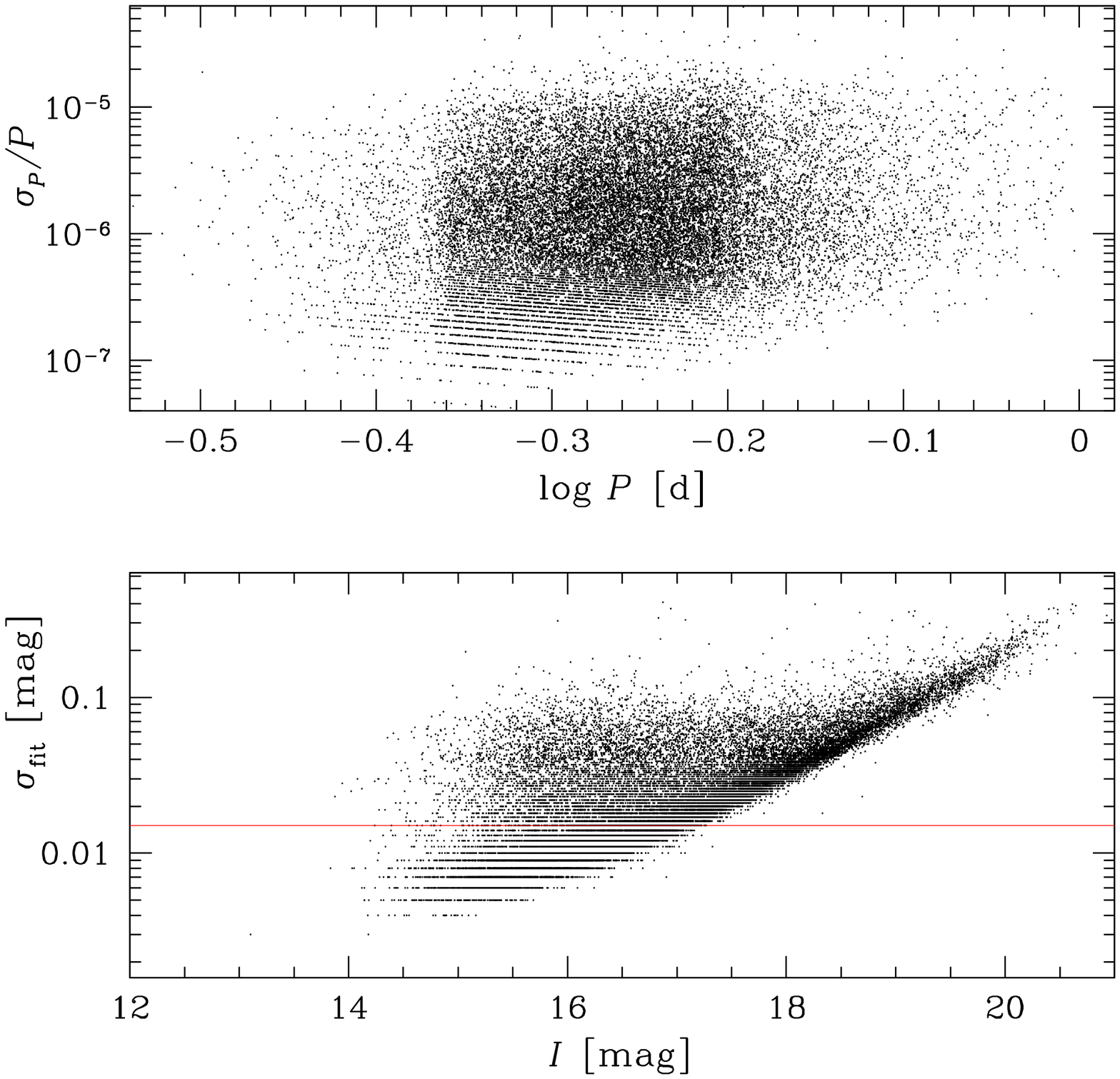}
\caption{
Period uncertainty as a function of period (upper panel) and
scatter around the fit as a function of magnitude (lower panel) for all
OGLE bulge RRab variables. Stars with $\sigma_{\rm fit}<0.015$~mag
(below red line) are used in the analysis of the period--amplitude diagram.}
\label{fig:sigmas}
\medskip
\end{figure}

Fig.~\ref{fig:pampvi} presents two period--amplitude (or Bailey) diagrams
for OGLE RRab stars, separately for amplitudes in the $I$ band and $V$ band.
Surprisingly, the stars form two very close sequences, well resolved
at low amplitudes. This is seen even better if we plot only stars with
a small scatter around the fit to the $I$-band data, $\sigma_{\rm fit}<0.015$~mag
(Fig.~\ref{fig:pamp015}). Such stars constitute about 27\% of the original
sample of bulge RRab stars. Our interpretation of the period--amplitude diagram
is that the two prominent sequences represent two major old bulge populations.
We denote the sequence with the shorter periods as population A and the one
with the longer ones as population B. The observed separation is much larger
than the uncertainty of the period (see plots of $\sigma_{P}$ vs. log~$P$
and $\sigma_{\rm fit}$ vs. $I$ in Fig.~\ref{fig:sigmas}).

\begin{figure}
\centering
\includegraphics[width=8.5cm]{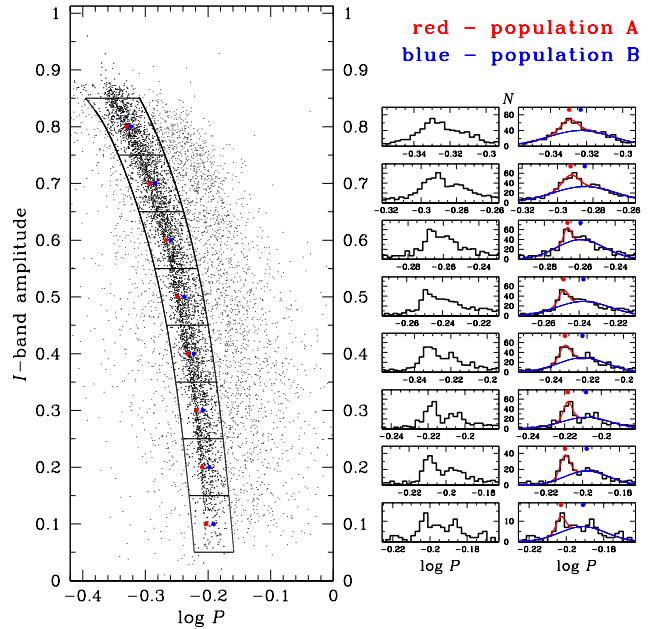}
\caption{
Period--amplitude diagram for bulge RRab variables with $\sigma_{\rm fit}<0.015$~mag
(on the left-hand side). The presence of two sequences is confirmed
on histograms (on the right-hand side) for stars in the marked boxes.
Results of the fitting of a sum of two Gaussians are presented in the histograms
in the right column. The two components/populations are shown with different
colors. The maximum values are marked with points above the histograms
and in the boxes in the diagram.}
\label{fig:pamppops}
\medskip
\end{figure}

In Fig.~\ref{fig:pamppops}, we analyze the period--amplitude diagram in a
quantitative way. First, we have to note that we tried to fit a second-order
polynomial (a parabola) to the general trend in the diagram, but it failed.
A fourth-order polynomial fits much better to the data for $I$-band amplitudes
$<0.85$~mag. We set a central line and nearly parallel boundaries on low
and high periods and divided the area between the lines into eight amplitude sectors
every 0.1~mag starting from 0.05~mag. In each sector, we counted stars in bins
along the trend. The obtained histograms are presented to the right of the
period--amplitude diagram in Fig.~\ref{fig:pamppops}. Two peaks are well separated
for amplitudes $\lesssim0.5$~mag. To resolve the two populations at higher amplitudes
and find general relations between them we fit a sum of two Gaussians
to the data. The fits are presented in the same figure. We can see that
the components are significantly different in width and their maxima get
closer with the increasing amplitude. An average shift between the populations
amounts to log~$P=0.0095(7)$. Population B is about four times more abundant than
population A. However, there are many stars beyond the analyzed area,
in particular with much shorter periods. They constitute about 40\% of the whole
sample. Therefore, we conclude that about 12\% of RR Lyrae stars belong to
population A, about 48\% to population B, and the remaining 40\% to other
populations.

By fitting fourth-order polynomials to the points of maxima in the period--amplitude
diagram, we find the following relations for the two populations:
\begin{eqnarray}
\label{equ:18}
a_{I,{\rm popA}}=-2400(\pm1100)~{\rm log}^4P- \nonumber\\
-2800(\pm1200)~{\rm log}^3P-1290(\pm500)~{\rm log}^2P- \nonumber\\
-264(\pm91)~{\rm log}P-20.1(\pm6.2),
\end{eqnarray}
\begin{eqnarray}
\label{equ:19}
a_{I,{\rm popB}}=-1740(\pm740)~{\rm log}^4P- \nonumber\\
-2040(\pm810)~{\rm log}^3P-910(\pm330)~{\rm log}^2P- \nonumber\\
-184(\pm58)~{\rm log}P-13.9(\pm3.8).
\end{eqnarray}

\begin{figure}
\centering
\includegraphics[width=8.5cm]{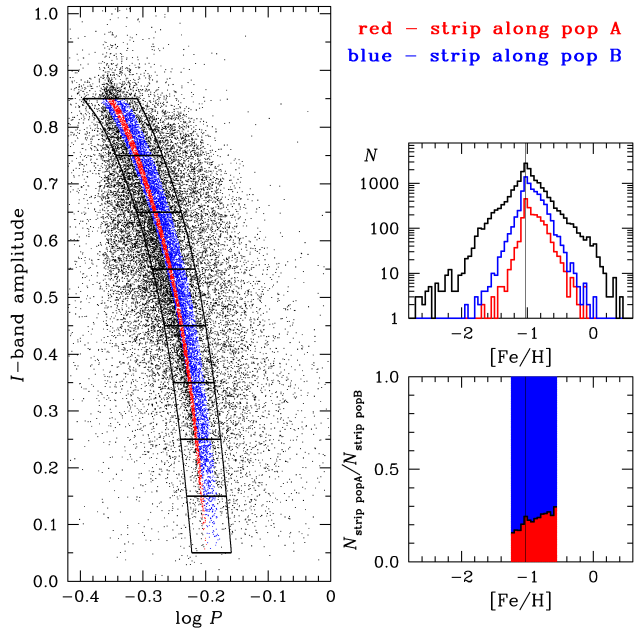}
\caption{
Left panel: period--amplitude diagram for all RRab variables with
location of $1\sigma$ strips along population A (in red) and population B
(in blue). Upper right: histograms of photometrically assessed metallicities
on the \cite{1995AcA....45..653J} scale for all RRab stars (black line),
stars from the strip along population A (in red), and stars from the strip
along population B (in blue). The bin size is 0.05~dex. The thin vertical line
marks the modal value of $\feh_{\rm J95}=-1.025$~dex. Lower right: normalized
proportion of stars from the two strips calculated in the same metallicity
bins if the total number of objects exceeds 100 per bin. Note that,
despite the same modal value, the distribution of stars from the strip
along population A is more skewed toward higher metallicities
than the one for stars from the strip along population B.}
\label{fig:pampfeh}
\medskip
\end{figure}

The two major old bulge populations must have very similar properties
since they practically overlap with each other in the period--amplitude
diagram. In particular, it is impossible to indicate which variable
belongs to which population in a strip along population A.
Theoretical (e.g., \citealt{1997ApJ...483..811B}) as well as empirical
investigations (e.g., \citealt{2015ApJ...798L..12F}) on RR Lyrae type
variables show that such a split in the Bailey diagram
is due to different iron abundance. Based on the present data
we can try to search for differences in the metallicity distributions
among groups of stars selected in the period--amplitude diagram.
In Fig.~\ref{fig:pampfeh}, we compare metallicity distributions for stars
from a $1\sigma$ strip along population A with a $1\sigma$ strip along
population B. Each strip is actually a mix of populations. The strip along
population B contains about 20\% of stars from population A, while the
strip along population A contains nearly 50\% of stars from population B.
Nevertheless, the difference in metallicity distributions is noticable---the
one along population A is more skewed toward higher metallicities.
In both cases the maximum value is at $\feh_{\rm J95}=-1.025$~dex,
but medians differ by 0.017~dex: $-0.947$~dex for the strip along
population A and $-0.964$~dex for the one along population B.

\begin{figure}
\centering
\includegraphics[width=8.5cm]{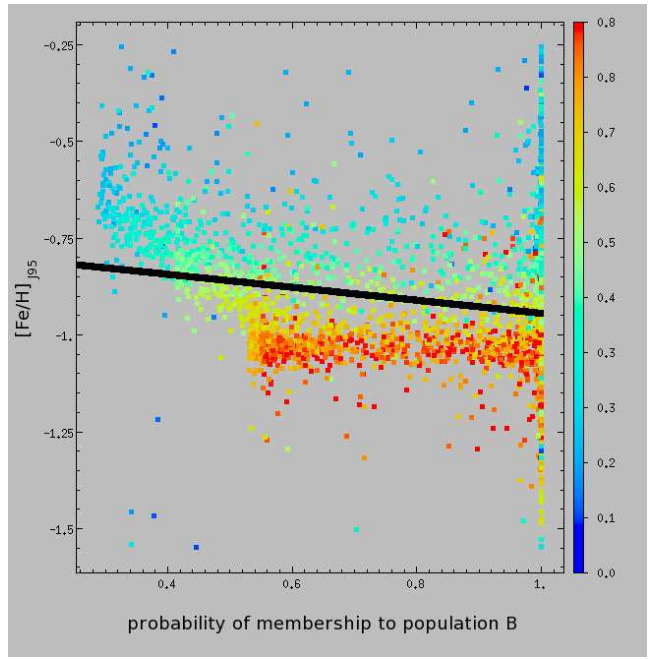}
\caption{
Metallicity distribution in a function of probability that a star
belongs to population B. Colors code the $I$-band amplitude.
Inclined black straight line fitted to the data indicates a difference
in metallicity of 0.17~dex for the probability from 1 to zero.
Population B is clearly more metal-poor than population A
and probably by this value.}
\label{fig:fehvsprob}
\medskip
\end{figure}

In Fig.~\ref{fig:fehvsprob}, we try to assess the metallicity
difference in another way. For each star we calculate the probability
of membership in population B and locate the stars in a
probability--metallicity diagram. To the formed distribution we fit
a straight line. It has $\feh_{\rm J95}=-0.94$~dex at the end
where probability reaches 1 and $\feh_{\rm J95}=-0.77$~dex
where probability is zero. It is clear again that, on average, stars from
population A are more rich in metals than stars from population B.
The difference of 0.17~dex may be overestimated, but it seems to
be a more realistic value than the 0.02~dex derived from the medians.

\section{Conclusions}
\label{sec:conc}

The recently released collection of 27,258 fundamental-mode RRab
type variable stars detected toward the Galactic bulge
(\citealt{2014AcA....64..177S}) by the OGLE survey has been used
to study the structure of the old bulge and its properties.
The present sample of bulge RRab stars includes newly discovered
variables in OGLE-IV and is about twice as larger as
the one analyzed in \cite{2012ApJ...750..169P}.

After initial cleaning of the sample from likely globular cluster
members and foreground and background objects (mainly from the Sgr
dSph galaxy), we investigated the surface density distribution of the
bulge RRab stars on the sky. We demonstrated that constant density lines
are ellipses flattened along the Galactic equator with a flattening
of $f=0.33\pm0.03$. We also found that the density profile can be
described by a single power law at observed Galactocentric distances
between 0.2 and 2.8~kpc for the obtained in this work distance to
the GC of $R_0=8.27\pm0.01{\rm(stat)}\pm0.40{\rm(sys)}$~kpc
that was obtained in this work.

The analysis of the spatial distribution of RR Lyrae stars shows
that they form a triaxial ellipsoid with a major axis located
in the Galactic plane and inclined at an angle of $i=20\degr\pm3\degr$
to the Sun--GC line of sight. This angle is close to the orientation
angle of the Galactic bar ($\sim30\degr$) traced by intermediate-age
RC giants. For the RR Lyrae ellipsoid, we determined the scale-length
ratio of the major axis to the minor axis in the plane, and to the
axis vertical to the plane $1:0.49(2):0.39(2)$. Such a spatial distribution
is consistent with the fact that RR Lyrae stars are in the same gravitational
potential together with the more massive Galactic bar. The result presented
here confirms a conclusion stated in \cite{2012ApJ...750..169P} after
a comparison of the brightness between the bulge RR Lyrae stars and RC
giants in a function of Galactic coordinates, illustrated in their
Figs.~12--14. The tilt angle of $12\fdg5\pm0\fdg5$ obtained by
\cite{2013ApJ...776L..19D} seems to be too small and the metal-poor
bulge population indeed trace closely the barred distribution of
RC stars. However, we cannot rule out a possibility that the geometric
properties of the old bulge are different in its central part.

From the parameters of the $I$-band light curves we assessed metallicities
of the bulge RRab stars on the \cite{1995AcA....45..653J} scale. We confirmed
our earlier finding (\citealt{2012ApJ...750..169P}) that the metallicity
distribution is sharply peaked at $\feh_{\rm J95}=-1.02$~dex, with a
dispersion of 0.25~dex. We also confirmed a very weak metallicity gradient
with Galactocentric distance recently noted in \cite{2014A&A...571A..59S}
based on the OGLE-III data. The analysis of the present sample gives
a value of $-0.018(3)$~dex~kpc$^{-1}$ and indicates that the shape
of the distribution does not change with the distance from the center.

In this work, for the first time, we pay attention to two very close
sequences formed by RRab stars in the period--amplitude diagram.
We interpret them as the presence of two major old bulge populations,
A and B, which differ in their iron content. The mean observed
shift between the two populations is 0.0095(7) in ${\rm log}~P$,
while an estimated difference in metallicity is about 0.17~dex or lower.
We found that population A is about four times less abundant
than the slightly more metal-poor population B. By counting stars
we have assessed that population A contains about 12\% of the bulge
RR Lyr stars, populations B about 48\%, and other old bulge
populations the remaining 40\%. The existence of multiple old populations
suggests that the early formation of the Milky Way bulge happened
via mergers (\citealt{2004ARA&A..42..603K,2015MNRAS.446.2837S}).

The large sample of RR Lyrae stars detected in the optical regime
by the OGLE survey has allowed us to study the 3D structure of the old
bulge as close as $\sim$0.3~kpc from the GC. Soon the ongoing
near-IR VVV survey should complete this picture for the most central
regions. However, there will still be many questions to answer,
in particular, on the assembly history and dynamics of the old
bulge and its relation to the Galactic halo. Spectroscopic
follow-up observations conducted by surveys such as APOGEE
(\citealt{2012ApJ...755L..25N}) and ARGOS (\citealt{2013MNRAS.428.3660F})
may help us to find answers to these key issues.


\acknowledgments

The OGLE project has received funding from the National Science
Centre, Poland, grant MAESTRO 2014/14/A/ST9/00121 to AU.
This work has been also supported by the Polish Ministry of Sciences
and Higher Education grants No. IP2012 005672 under the Iuventus Plus
program to PP and No. IdP2012 000162 under the Ideas Plus program to IS.


\appendix

\section{Determination of the $c/a$ ratio}
\label{sec:c2aratio}

\begin{figure}
\centering
\includegraphics[width=8.5cm]{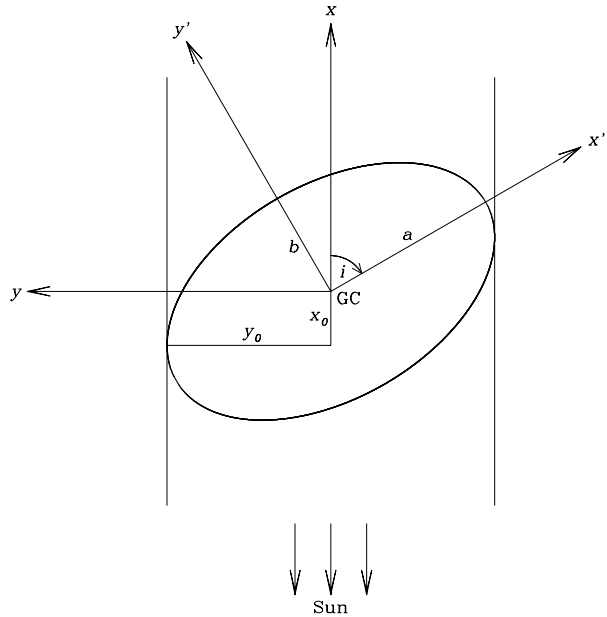}
\caption{
Cross section of an ellipsoid with the semiaxes $a$ and $b$ in the
Galactic plane and inclined at an angle $i$ to the Sun--GC line of sight.}
\label{fig:ellipse}
\medskip
\end{figure}

If we assume that the viewing angle at which we observe the RR Lyr
ellipsoid is negligible, the $c/a$ ratio can be estimated from simple
geometric considerations. For an ellipsoid with the axes $a$
and $b$ in the $xy$ (Galactic) plane, and an axis $c$ perpendicular
to this plane, we can write:
\begin{equation}
\label{equ:A1}
\frac{c}{a} = \frac{c}{y_0}\frac{y_0}{a},
\end{equation}
where $y_0$ is the projected half-width in the plane, as shown in the
cross section in Fig.~\ref{fig:ellipse}. In our case $c/y_0$ roughly equals
to $r_b/r_l=1-f$. The proportion $y_0/a$ can be determined by drawing
tangent lines parallel to the inclined ellipse which is a result
of the plane cross section of the ellipsoid. The standard equation for
the ellipse in the $x^{\prime} y^{\prime}$ coordinate system is
\begin{equation}
\label{equ:A2}
\frac{x'^2}{a^2} + \frac{y'^2}{b^2} = 1.
\end{equation}
The $x^{\prime} y^{\prime}$ system is rotated with respect to the $xy$ system
through the clockwise angle $i$, hence
\begin{equation}
\label{equ:A3}
\frac{(x{\rm cos}i+y{\rm sin}i)^2}{a^2} + \frac{(x{\rm sin}i-y{\rm cos}i)^2}{b^2} = 1.
\end{equation}
Expanding the binomial squares and collecting like terms leads
to the following quadratic equation:
\begin{equation}
\label{equ:A4}
Ay^2 + By + C = 0,
\end{equation}
where
$$
A = \frac{{\rm sin}^2i}{a^2}+\frac{{\rm cos}^2i}{b^2},
$$
$$
B = -2x\Big(\frac{1}{b^2}-\frac{1}{a^2} \Big){\rm sin}i{\rm cos}i,
$$
$$
C = \Bigg(\frac{{\rm cos}^2i}{a^2}+\frac{{\rm sin}^2i}{b^2} \Bigg)x^2-1.
$$
The discriminant of the above quadratic equation
\begin{equation}
\label{equ:A5}
\Delta = B^2-4AC = \frac{4}{a^2b^2}(b^2{\rm sin}^2i+a^2{\rm cos}^2i-x^2).
\end{equation}
The maximum of the root
\begin{equation}
\label{equ:A6}
y = \frac{-B+\sqrt{\Delta}}{2A}
\end{equation}
is the searched half-width $y_0$. From the derivative $dy/dx=0$, we find
\begin{equation}
\label{equ:A7}
\frac{x_0}{a}=\Bigg(1-\Big(\frac{b}{a}\Big)^2\Bigg){\rm sin}i{\rm cos}i\sqrt\frac{\Big(\frac{b}{a}\Big)^2{\rm sin}^2i+{\rm cos}^2i}{\Big(\frac{b}{a}\Big)^2+\Big(1-\Big(\frac{b}{a}\Big)^2\Big)^2{\rm sin^2}i{\rm cos}^2i}.
\end{equation}
Finally, we obtain
\begin{equation}
\label{equ:A8}
\frac{y_0}{a}=\frac{\Bigg(1-\Big(\frac{b}{a}\Big)^2\Bigg)\frac{x_0}{a}{\rm sin}i{\rm cos}i+\frac{b}{a}\sqrt{\Big(\frac{b}{a}\Big)^2{\rm sin}^2i+{\rm cos}^2i-\Big(\frac{x_0}{a}\Big)^2}}{\Big(\frac{b}{a}\Big)^2{\rm sin}^2i+{\rm cos}^2i}.
\end{equation}

\end{document}